\documentclass[aps,prl,superscriptaddress,twocolumn,showpacs,preprintnumbers,nofootinbib]{revtex4-1}
\usepackage{amsfonts}
\usepackage{amssymb,amsmath}
\usepackage{mathrsfs}
\usepackage{amsthm}
\usepackage{newlfont}
\usepackage{graphicx}

\newcommand{\usb}{\affiliation{Departamento de F\'{\i}sica, Secci\'{o}n de Fen\'{o}menos \'{O}pticos, Universidad Sim\'{o}n Bol\'{\i}var,Apartado Postal 89000, Caracas 1080-A, Venezuela.}}
\newcommand{\ivic}{\affiliation{Centro de F\'{\i}sica, Instituto Venezolano de Investigaciones Cient\'{\i}ficas, Apartado 20632 Caracas 1020-A, Venezuela.}}

\begin{document}
\begin{flushright} ${}$\\[-40pt] $\scriptstyle \mathrm SB/F/437-14$ \\[0pt]
\end{flushright}
\title{Belinfante-Rosenfeld tensor and the inertia principle }
\author{Rodrigo Medina}\email[]{rmedina@ivic.gob.ve}\ivic
\author{J Stephany}\email[]{stephany@usb.ve}\usb
\date{\today}

\begin{abstract}
In  a recent letter we show that for an isolated system with a non symmetric energy momentum tensor the usual forms of the center of mass motion theorem are not valid. This was illustrated with a particular configuration of a magnet and a point charge for which it was shown that what is usually regarded as the center of mass of the system does not remain stationary even if the system is isolated. In a subsequent work we demonstrated that the violation of the center of mass motion theorem for isolated systems with spin is a direct consequence of the conservation of total angular momentum. We also show that there exists a generalized center of mass and spin which moves with constant velocity. In this letter we show that this center of mass and spin corresponds to the center of mass defined by  the Belinfante-Rosenfeld tensor. We also show that, if the  spin density instead of being of microscopic origin appears by a scaling process,  the macroscopic Belinfante-Rosenfeld tensor emerges from the average of the 
microscopic energy-momentum tensor as the true macroscopic energy momentum tensor. This implies that in general spin has to be considered as a source of the  gravitational field in general relativity. 
\end{abstract}
\maketitle

\bigskip

\section{Introduction}
There has been always some ambiguity in the literature when referring  to the energy momentum tensor of a relativistic system. In general relativity it is the source of Einstein's equations necessarily a symmetric tensor which bears this name. In the Lagrangian field theoretical set up one finds the canonical energy-momentum tensor which emerges from Noethers's theorem and  the symmetric Belinfante-Rosenfeld \cite{BelF1940,Rosenfeld1940} tensor which is a combination of the canonical tensor and the derivatives of the spin density. Finally for an interacting field-matter-energy  system there should exist what could be named the kinetic energy-momentum tensor which accounts for the energy and momentum exchange determined by the density of force acting upon it. This object is not necessarily equal to any of the two above even if a Lagrangian description is available. Although most of the ambiguity is a matter of words and the true meaning of the energy-momentum tensor being used is almost always fixed by the 
context some confusion found its way in  a 
particular aspect of the description of relativistic systems, the so called center of mass motion theorem (CMMT). In non relativistic mechanics this theorem states that the center of mass of an isolated system moves with constant velocity and is a consequence of Newton's third law. In relativistic physics a CMMT theorem may be  derived from the conservation of orbital angular momentum but requires the energy momentum tensor to be symmetric. It comes in two versions. The first (CMMT-1) states that the center of mass of an isolated system moves with constant velocity. The second (CMMT-2) states that the total momentum of an isolated system equals the mass (energy divided by $c^2$) times the
velocity of the center of mass. The second version implies the first. 

In a recent letter \cite{MedandSa}  we made the simple observation that when the energy-momentum tensor is non-symmetric  none of these forms of the theorem  hold. Instead a third version (CMMT-3) of the theorem which relates the energy current to the velocity of the center of mass is shown to be valid for any isolated system. Defining the center of mass by
\begin{equation}
\label{CenterMass}
X^i_T = \frac{1}{U}\int x^i T^{00}\,dV\ \ .
\end{equation} 
with $U=\int T^{00}\,dV$, the CMMT-3 states that,
\begin{equation}
\label{CenterMass-velocity}
\dot{X}^i_T = \frac{c}{U}\int T^{0i}\,dV\ \ .
\end{equation}
Note that what appears in (\ref{CenterMass-velocity}) is the energy current density, not the momentum density $c^{-1}T^{i0}$. In the same paper we show how Maxwell equations and the use of Lorentz force determine the violation of CMMT-1 and CMMT-2 in a particular charge-magnet system.  This allows us to infer that the energy momentum tensor of this system should be  non-symmetric. In fact in \cite{MedandSb} we deduce the appropriate non-symmetric kinetic energy-momentum tensor for the electromagnetic field in matter exploiting the consistency of the macroscopic Maxwell equations and the Lorentz force. The correct form  of the macroscopic Lorentz force and of the electromagnetic kinetic energy-momentum  was also corroborated by an independent computation taking the averages over the microscopic equations in Ref.\cite{MedandSc}. The obtained energy-momentum tensor  coincides neither with Abraham's \cite{Abraham1910} nor with Minkowski's \cite{Minkowski1909} expressions but vindicates the   density of momentum 
of the electromagnetic field in matter proposed by Minkowski as the one which assures momentum conservation.

The violation of the CMMT has important  implications to the fundamental structure of  relativistic mechanics. The inertial reference frames necessary to discuss the inertia principle are defined as those for which isolated systems satisfy the CMMT. To rescue the inertia principle, one has either to restrict the category of systems used to define the inertial reference frames or one has to find a substitute to the CMMT. With this in mind, in Ref. \cite{MedandStef} we show that the possible violation of CMMT in systems with spin is consequence of the conservation of total angular momentum. To prove this we show that it is a generalized quantity that we call the center of mass and spin that moves with constant velocity. Below we show that the center of mass and spin corresponds to the center of mass defined in terms of the Belinfante-Rosenfeld tensor.

The Belinfante-Rosenfeld tensor is a combination of the energy momentum tensor and the derivatives of the spin density tensor. For relativistic isolated systems it is symmetric and conserved. It was  invoked, first  by Coleman and van Vleck \cite{CvV1968} and then by many others to claim that  a  symmetric energy momentum tensor is always available. The conviction that  a  symmetric energy momentum tensor is always available lead  Shockley and James \cite{SJ1967,Pfeifer2007} to try to compensate the  evidence of the violation of CMMT-1 and CMMT-2 in Maxwell-Lorentz electromagnetism by  proposing the hidden momentum hypothesis. For the same reasons other authors engaged in a search for alternatives to the Lorentz force density \cite{Pfeifer2007}. Both options depart from established physics.  On the other hand it  straightforward to see that  the center of mass like object defined with the Belinfante-Rosenfeld tensor satisfies CMMT-1 and CMMT-2. How could this be made consistent with the results mentioned 
above which exhibit a system with a non inertial center of mass? Here is where the ambiguity mentioned at the beginning  enters again. The very simple answer  which has been in the light spot but unnoticed for 
more than 70  years is  that the  center of mass defined with the true kinetic energy-momentum tensor or with the Belinfante-Rosenfeld tensor are not the same object. The object defined with the Belinfante-Rosenfeld tensor includes a contribution of the spin density which has to be taken into account in the dynamics. As shown below it corresponds to the center of mass and spin defined in  \cite{MedandStef}. The objective of  this letter is to clarify this point and discuss how spin enters in this improved Center of Mass and Spin Motion Theorem (CMSMT).

\section{Belinfante-Rosenfeld construction}
Consider a relativistic isolated system described by a non-symmetric energy momentum tensor $T^{\mu\nu}$ which is conserved  $\partial_\nu T^{\mu\nu} =0$ and a spin current density  $S^{\mu\nu\alpha}=-S^{\nu\mu\alpha}$. The orbital angular momentum current is defined by $L^{\mu\nu\alpha}=x^\mu T^{\nu\alpha}-x^\nu T^{\mu\alpha}$. The total angular momentum current $J^{\mu\nu\alpha}=L^{\mu\nu\alpha}+S^{\mu\nu\alpha}$ satisfy, $\partial_\alpha J^{\mu\nu\alpha} =0$ so that the total angular momentum $J^{\mu\nu}=L^{\mu\nu}+S^{\mu\nu}$ with $L^{\mu\nu}=c^{-1}\int L^{\mu\nu 0}dV$ and $S^{\mu\nu}=c^{-1}\int S^{\mu\nu 0}dV$ is conserved. 
\begin{equation}
\label{CTAM}
 \frac{d}{dt} J^{\mu\nu}=\frac{d}{dt} (L^{\mu\nu}+S^{\mu\nu})=0
\end{equation}
Then we have the equations 
\begin{equation}
\label{orbital-eq}
\partial_\alpha L^{\mu\nu\alpha} = T^{\nu\mu}-T^{\mu\nu} \ \ .
\end{equation}
and
\begin{equation}
\partial_\alpha S^{\mu\nu\alpha} = -T^{\nu\mu}+T^{\mu\nu} \ \ .
\end{equation}
This equations hold when  dealing with  canonical objects obtained via Noether's theorem from a Lagrangian but also when working with the kinetic objects associated to a matter distribution \cite{Papapetrou1949,MedandSd}. 
Defining
\begin{equation}
 K^{\mu\nu\alpha}=\frac{1}{2}[ S^{\mu\alpha\nu}+S^{\nu\alpha\mu}-S^{\mu\nu\alpha}]
\end{equation}
the Belinfante-Rosenfeld tensor is given by
\begin{equation}
\label{BelRos}
 \Theta^{\mu\nu}=T^{\mu\nu}+K^{\mu\nu}\ \  ,\ \ K^{\mu\nu}=\partial_\alpha K^{\mu\nu\alpha}
\end{equation}
The Belinfante-Rosenfeld tensor is symmetric and conserved, $\partial_\nu \Theta^{\mu\nu} =0$. Moreover the conserved charges computed with this tensor are equal to the  four momentum of the system,
\begin{equation}
 \frac{1}{c}\int \Theta^{\mu 0}dV=\frac{1}{c}\int T^{\mu 0}\, dV=P^\mu
\end{equation}
so that in particular the total energy is $U=\int \Theta^{00}\,dV$. The total angular momentum may be written in the form 
\begin{equation}
 J^{\mu\nu}=c^{-1}\int J^{\mu\nu 0}\, dV=c^{-1}\int (x^\mu \Theta^{\nu 0}-x^\nu \Theta^{\mu 0})\, dV. 
\end{equation}
The meaning of this equation is only that the spin  is incorporated to the modified orbital angular computed with the Belinfante-Rosenfeld tensor.

\section{The CMSMT theorem}
Let us define the quantity corresponding to (\ref{CenterMass}) using Belinfante-Rosenfeld tensor $\Theta^{\mu\nu}$ and name it the center of mass and spin,
\begin{equation}
\label{CenterMassSpin}
X^i_\Theta = \frac{1}{U}\int x^i \Theta^{00}\,dV\ \ .
\end{equation} 
Then it is straightforward to show \cite{MedandSa} that
\begin{equation}
\label{CenterMassSpin-velocity}
\dot{X}^i_\Theta = \frac{c}{U}\int \Theta^{0i}\,dV= \frac{c^2 P^i}{U}\ \ ,
\end{equation}
which resembles the usual center of mass motion theorem. But writing, 
\begin{equation}
\label{CMB}
X^i_S= \frac{1}{U}\int x^i \partial_\alpha K^{00\alpha}\,dV= -\frac{c}{U}S^{0i}                                                                 
\end{equation}
we have
\begin{equation}
\label{CenterMassSpinsplit}
{X}^i_\Theta = {X}^i_T+{X}^i_S .
\end{equation}
which shows that in (\ref{CenterMassSpin-velocity}) there is a term which is not related to mass, energy or momentum but only to spin. As mentioned, (\ref{CenterMassSpinsplit}) is the center of mass and spin defined in \cite{MedandStef}. The fact that $X_\Theta$ is indeed inertial follows  directly from the conservation of the total angular momentum\cite{MedandStef}. To see this we note that using  (\ref{CTAM}),
\begin{eqnarray}
\label{CMSMT}
\frac{d}{dt}X^i_S&=&\frac{c}{U} \frac{d}{dt} L^{0i}= \frac{1}{U}\frac{d}{dt}\int \big[x^0 T^{i0}-x^iT^{00}\big]dv\nonumber\\
&=&\frac{c^2 P^i}{U}-\frac{d}{dt}X^i_T\ .
\end{eqnarray}

If the spin current density $S^{\mu\nu\alpha}$ considered above  comes from the quantum mechanical spin of the atoms and particles in the matter distribution,  the  energy-momentum tensor which describes the dynamics at the scale where  $S^{\mu\nu\alpha}$ is relevant would be non-symmetric \cite{Papapetrou1949,MedandSa} . This is the case for the system considered in \cite{MedandSa}
formed by a toroidal magnet with a charge in its center initially at rest. The spin density in this situation has contributions from the magnetization and from the electromagnetic field.  When the magnet is heated and the magnetization disappears Maxwell equations impose that  $X^i_T$ is not stationary \cite{MedandSa}. But $X^i_\Theta$ remains at rest. The field carries away the energy like contribution defined through $K^{\mu\nu}$ in (\ref{BelRos}) and (\ref{CenterMassSpin}) and $X^i_S$ moves in the opposite direction that $X^i_T$. The actual spin dynamics of this system is complicated and the spin densities have to be defined and worked with much care, but the general results (\ref{CenterMassSpin-velocity}),(\ref{CenterMassSpinsplit}) are independent of the details.

It is interesting to note that the lack of symmetry of the kinetic energy-momentum tensor is another way to display the quantum nature of magnetism. If one tries to represent the magnetic moments with  microscopic  currents, for the whole description it would be  necessary to consider also the stresses on the conductors and then, and one ends up with a symmetric total energy-momentum tensor. It is necessary to introduce a true spin density to have a non-symmetric energy momentum tensor at the microscopic scale. 

\section{Spin from average}
To complete the picture let us see what happens if the spin density is not fundamental but comes from averaging in space at some scale. We have in mind for example an application to general relativity where  the rotational energy  of compact objects or galaxies \cite{MedandSd} is to be included.  Consider a system with a conserved symmetric microscopic energy-momentum tensor $T^{\mu\nu}_{\mathrm mic}$ and a conserved total orbital angular momentum $J^{\mu\nu\alpha}$ corresponding to a situation in which there is no microscopic spin,
\begin{equation}
 J^{\mu\nu\alpha}_{\mathrm mic}=x^\mu T^{\nu\alpha}_{\mathrm mic}-x^\nu T^{\mu\alpha}_{\mathrm mic} \ .
\end{equation}
We want the spin density emerge from an average process at some scale so we divide the energy-momentum tensor in two portions not necessarily symmetric $T^{\mu\nu}_{\mathrm M}$ and $T^{\mu\nu}_R$, with the latter corresponding to the rotational energy at that scale
\begin{equation}
T^{\mu\nu}_{\mathrm mic}=T^{\mu\nu}_{\mathrm M} + T^{\mu\nu}_{\mathrm R} \ .
\end{equation}
Since  $T^{\mu\nu}_{\mathrm mic}$ is symmetric we have
\begin{equation}
 T^{\mu\nu}_{\mathrm M}-T^{\nu\mu}_{\mathrm M}=-(T^{\mu\nu}_{\mathrm R}-T^{\nu\mu}_{\mathrm R})
\end{equation}
The microscopic angular momentum divides also in two portions 
\begin{equation}
 J^{\mu\nu\alpha}_{\mathrm mic}= L^{\mu\nu\alpha}_{\mathrm mic}
 =L^{\mu\nu\alpha}_{\mathrm M}+L^{\mu\nu\alpha}_{\mathrm R} \ ,
\end{equation}
which are given by 
\begin{equation}
 L^{\mu\nu\alpha}_{\mathrm M}=x^\mu T^{\nu\alpha}_{\mathrm M}-x^\nu T^{\mu\alpha}_{\mathrm M} \ .
\end{equation}
and
\begin{equation}
 L^{\mu\nu\alpha}_{\mathrm R}=x^\mu T^{\nu\alpha}_{\mathrm R}-x^\nu T^{\mu\alpha}_{\mathrm R} \ .
\end{equation}

To  define macroscopic quantities we take averages over small regions of space-time in the form,
\begin{equation}
\label{average}
\langle A(x)\rangle = \int A(x^\prime)W(x-x^\prime)\,dx^\prime\ ,
\end{equation}
where $W(x)$ is a smooth function that is essentially constant inside a region
of size $R$ and that vanishes outside
\begin{eqnarray}
W(x)\ge 0\ ,&\qquad& |x^\mu|\rangle R \Longrightarrow W(x)=0\ ,\nonumber\\
\qquad \int W(x)\,dx &=&1\ .
\end{eqnarray}
Inside the region $W(x)\approx \langle W\rangle$.

At the scale $R$ the microscopic fluctuations are washed out,
and  therefore all the products of averages and fluctuations are
negligible. That is, if $\delta A= A-\langle A\rangle$ then
$\langle \langle B\rangle\delta A\rangle\approx 0$ and also
 $\langle\langle B\rangle\rangle \approx \langle B\rangle$.
With  these conditions it follows that $\partial_\nu\langle A\rangle =\langle \partial_\nu A\rangle $.

In particular we define the macroscopic energy momentum tensor
\begin{equation}
 {\bar T}^{\mu\nu}=\langle T^{\mu\nu}_{\mathrm M}\rangle  .
\end{equation}
which we  suppose remains conserved $\partial_\nu {\bar T}^{\mu\nu}=0$. We can write for the averaged energy-momentum
\begin{equation}
 \langle T^{\mu\nu}_{\mathrm mic}\rangle ={\bar T}^{\mu\nu}+\langle T^{\mu\nu}_{\mathrm R}\rangle  \ .
\end{equation}
Since $\langle T^{\mu\nu}_{\mathrm mic}\rangle $ is conserved it follows that $\langle T^{\mu\nu}_{\mathrm R}\rangle $ is also conserved.
 
The averaged total angular momentum should be equal to the sum of the  macroscopic orbital angular momentum ${\bar L}^{\mu\nu\alpha}=x^\mu {\bar T}^{\nu\alpha}-x^\mu {\bar T}^{\mu\alpha}$ and the emerging spin,
\begin{equation}
\label{AvTAM}
 \bar J^{\mu\nu\alpha}=\langle J^{\mu\nu\alpha}_{\mathrm mic}\rangle ={\bar L}^{\mu\nu\alpha}+{\bar S}^{\mu\nu\alpha}
\end{equation}
and being a conserved quantity is necessary that,
\begin{equation}
 \partial_\alpha {\bar L}^{\mu\nu\alpha}={\bar T}^{\nu\mu}-{\bar T}^{\mu\nu}=-\partial_\alpha {\bar S}^{\mu\nu\alpha}
\end{equation}
On the other hand we have
\begin{eqnarray}
\langle J^{\mu\nu\alpha}_{\mathrm mic}\rangle =\langle L^{\mu\nu\alpha}_{\mathrm M}\rangle +\langle  L^{\mu\nu\alpha}_{\mathrm R}\rangle 
\end{eqnarray}
Writing,
\begin{equation}
 \langle  L^{\mu\nu\alpha}_{\mathrm M}\rangle ={\bar L}^{\mu\nu\alpha}+ \delta L^{\mu\nu\alpha}_{\mathrm M}
\end{equation}
it should be that the spin density is given by
\begin{equation}
\label{AvSpin}
{\bar S}^{\mu\nu\alpha}=\langle L^{\mu\nu\alpha}_{\mathrm R}\rangle +\delta L^{\mu\nu\alpha}_{\mathrm M}
\end{equation}
The true energy momentum distribution of the system at the macroscopic scale is given by $\langle T^{\mu\nu}_{\mathrm mic}\rangle $ which is a symmetric tensor and should be equal to the macroscopic Belinfante-Rosenfeld tensor which is the available symmetric object. In terms of the macroscopic quantities,
\begin{equation}
 \langle T^{\mu\nu}_{\mathrm mic}\rangle ={\bar T}^{\mu\nu}+ \bar K^{\mu\nu} 
\end{equation}
with
\begin{equation}
\bar  K^{\mu\nu}=\frac{1}{2} \partial_\alpha [ {\bar S}^{\mu\alpha\nu}+{\bar S}^{\nu\alpha\mu}-{\bar S}^{\mu\nu\alpha}]
\end{equation}
Then it is necessary that 
\begin{equation}
\label{AvK}
 \bar K^{\mu\nu}= \langle T^{\mu\nu}_{\mathrm R}\rangle 
\end{equation}
Equations (\ref{AvTAM}), (\ref{AvSpin}) and (\ref{AvK}) define the the macroscopic spin density.
The Belinfante-Rosenfeld tensor in this case is the true energy momentum tensor. Here spin is just another way of referring to the microscopic orbital angular momentum.  From the point of view of CMSMT this is necessary. The inertial center of mass described by $T^{\mu\nu}_{\mathrm mic}$ should be related after the averaging process with an inertial object which can be no other than  the center of mass and spin defined by the macroscopic Belinfante-Rosenfeld tensor. 

\section{Conclusion}
In this letter we show how taking at face value the meaning of Belinfante-Rosenfeld tensor one is led to the striking prediction that it is the space-time evolution of the spin density that restores the Newtonian  inertial principle  which, nevertheless requires the  generalized notion of center of mass and spin. This gives an interesting shift to the discussion presented  in \cite{MedandSa} which solved some  inconsistencies in classical electrodynamics which troubled physicists for many years but affects the inertia principle. From the discussion presented here is clear that the inertial reference frames should be defined as those for which isolated systems satisfy the CMSMT. In particular the result in this letter presents a further argument to abandon the hidden-momentum hypothesis that  in the last decades has been discussed in order to explain the failure of the CMMT in some  electromagnetic systems. In fact it highlights exactly where the analysis which led to that hypothesis went wrong and why.

The CMSMT presented above should be  a stimulus to retake the investigation on the dynamics of spin in this context. Since the early remarks of Gordon \cite{Gordon1928}, the kinematic implications of spin in field theory have been present in the work of many people. But a detailed account of how spin is exchanged between for example, the electromagnetic field and a material media is still lacking. In a treatment of this problem the distinction between the canonical and the kinematic spin density for the electromagnetic field will be specially relevant. 

The kinematical implications of  a space traveling spin density have a potential impact in particle physics where  additional restrictions  imposed by the movement of the center of mass and spin may in principle  appear on the trajectories of the particles after a collision, on top of momentum and angular momentum conservation. This may be relevant in the analysis of collisions of polarized particles or for understanding better the contributions of the quark-gluon sea for  the spin of hadrons. 

We also show by an explicit computation, that if instead of being of microscopic origin, the  spin density appears by a scaling process, it is the macroscopic Belinfante-Rosenfeld tensor which emerges from the average of the microscopic energy-momentum tensor.   This result  gives an interesting clue on why, as is  well known  \cite{Rosenfeld1940}, in the flat space-time limit the source of Einstein equations coupled to a field theory is not the energy-momentum tensor but the Belinfante-Rosenfeld tensor. An unavoidable consequence of the discussion is to note that,  besides energy and momentum  also spin is a source of the gravitational field. Pursuing this reasoning may be of interest not only to discuss compact systems, like a magnetized neutron star but principally to work out its consequences in cosmological models where the spin density may take care of part of the dark matter\cite{MedandSd}.  

We  hope that the  analysis presented in this letter could  also have a pedagogical impact in promoting to make a better distinction of the roles of the different objects being  called energy-momentum tensor in the literature and to disclose its relations. Whenever only global quantities as the momentum or the energy are involved it is true that the Belinfante-Rosenfeld tensor may be used instead of the kinetic energy-momentum tensor (or the canonical energy momentum tensor in the quantum theory). But when the local dynamics is relevant or even when it is  necessary to discern between spin an orbital angular momentum this is not longer true.


\begin{thebibliography}{long}
\bibitem{BelF1940} F.~J.~Belinfante, Physica \textbf{VI}, 887 (1939).
\bibitem{Rosenfeld1940} L.~Rosenfeld, Mem. Acad. Roy. Belg. (Cl. Sciences) \textbf{18},
fasc. 6, 2-3 (1940).
\bibitem{MedandSa} Rodrigo Medina and J.~Stephany, {\it Violation of the center of mass theorem for systems with electromagnetic interaction}, Preprint. 
\bibitem{MedandSb} Rodrigo Medina and J.~Stephany, {\it The force density and the kinetic energy-momentum tensor of electromagnetic fields in matter}, Preprint.
\bibitem{MedandSc} Rodrigo Medina and J.~Stephany, {\it Electromagnetic fields in matter revisited}, Preprint. 
\bibitem{Abraham1910} M.~Abraham, Rend. Circ. Mat. Palermo \textbf{30}, 33 (1910).
\bibitem{Minkowski1909} H.~Minkowski, Nachr. Ges. Wiss. Gottingen, 53 (1909).
\bibitem{MedandStef}Rodrigo Medina and J.~Stephany, {\it An improved inertia principle}, arXive:1404.1590. 
\bibitem{SJ1967} W.~Shockley and R.P.~James, Phys.~Rev.~Lett. \textbf{18}, 876 (1967).
\bibitem{Pfeifer2007} Robert N.~C.~Pfeifer {\it et al}, Rev.~Mod.~Phys.
\textbf{79}, 1197--1216 (2007).
\bibitem{CvV1968} Sidney Coleman and J.H.~Van Vleck Phys.~Rev. \textbf{171}, 1370 (1968).
\bibitem{Papapetrou1949} A.~Papapetrou, Phil. Mag. \textbf{40}, 937--946 (1949).
\bibitem{MedandSd} Rodrigo Medina and J.~Stephany, {\it Spin densities and non-symmetric energy momentum tensors in general relativity}, Preprint. 
\bibitem{Gordon1928}W.~Gordon, Z.Phys \textbf{50}, 6306 (1928)
\end{thebibliography}
\end{document}